\documentclass[english,aps,pre,showpacs,reprint]{revtex4-1}
\usepackage[T1]{fontenc}
\usepackage[latin9]{inputenc}
\usepackage{float}
\usepackage{mathtools}
\usepackage{amsmath}
\usepackage{amssymb}
\usepackage{graphicx}
\usepackage{babel}

\begin{document}
	
\title{Adaptive Scales of Spatial Integration and Response Latencies in a Critically-Balanced Model of the Primary Visual Cortex}

\date{\today}

\author{Keith Hayton}
\author{Dimitrios Moirogiannis}
\author{Marcelo Magnasco}
\affiliation{Laboratory of Integrative Neuroscience, The Rockefeller University}

\begin{abstract}
The primary visual cortex (V1) integrates information over scales in visual space, which have been shown to vary, in an input-dependent manner, as a function of contrast and other visual parameters. Which algorithms the brain uses to achieve this feat are largely unknown and an open problem in visual neuroscience. We demonstrate that a simple dynamical mechanism can account for this contrast-dependent scale of integration in visuotopic space as well as connect this property to two other stimulus-dependent features of V1: extents of lateral integration on the cortical surface and response latencies. 
\end{abstract}

\maketitle

\section*{Introduction}
Stimuli in the natural world have quantitative characteristics that vary over staggering ranges. Our nervous system evolved to parse such widely-ranging stimuli, and research into how the nervous system can cope with such ranges has led to considerable advances in our understanding of neural circuitry. For example at the sensory transduction level, the physical magnitudes encoded into primary sensors, such as light intensity, sound pressure level and olfactant concentration, vary over exponentially-large ranges, leading to the Weber-Fechner law \cite{fechner1860elements}. As neuronal firing rates cannot vary over such large ranges, the encoding process must compress physical stimuli into the far more limited ranges of neural activity that represent them. These observations have stimulated a large amount of research into the mechanisms underlying {\em nonlinearly compression} of physical stimuli in the nervous system . Of relevance to our later discussion is the nonlinear compression of sound intensity in the early auditory pathways \cite{eguiluz2000essential, camalet2000auditory}, where it has been shown that poising the active cochlear elements on a Hopf bifurcation leads to cubic-root compression. 

\begin{figure}[bth]
	
		\begin{centering}
			\includegraphics[scale=0.4]{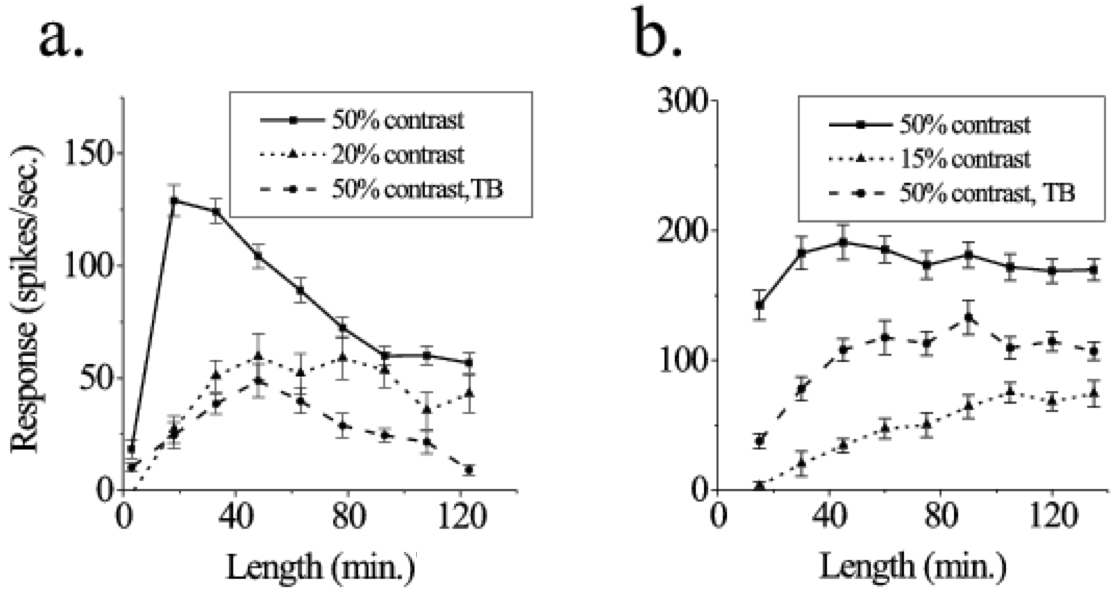}
			\par\end{centering}

	\caption{\label{fig1} \textbf{Reprinted from \cite{kapadia1999dynamics} with permission: Copyright (1999) National Academy of Sciences, U.S.A. Measurements of single-neuron responses in the V1 area of monkeys to optimally oriented bars of light of different lengths and contrasts.} Panels a and b are measurements from two distinct neurons. The units of length along the horizontal axis are in minutes of arc. The solid, dotted, and dashed curves represent bars of light of 50\% contrast, 15\% contrast, and 50\% contrast embedded in a textured background, respectively. The dashed curves are irrelevant to the focus of this paper.}
\end{figure}

But other characteristics besides the raw physical magnitude still vary hugely. The wide range of spatial extents and correlated linear structures present in visual scenery \cite{sigman2001common,field1987relations,ruderman1994statistics} leads to a more subtle problem, if we think of the visual areas as fundamentally limited by corresponding anatomical connectivity. Research into this problem has been focused on elucidating the nature of receptive fields of neurons in the primary visual cortex (V1) \cite{kapadia1995imporovement,sceniak1999contrast,zipser1996contextual,polat1998collinear,levitt1997contrast,kapadia1999dynamics}.  Studies have found that as the contrast of a stimulus is decreased, the receptive field \cite{kuffler1953discharge,hubel1962receptive} size or area of spatial summation in visual space increases (Fig \ref{fig1}) \cite{kapadia1999dynamics,sceniak1999contrast,deangelis1994length,deangelis1992organization}. As an example of contextual modulation of neuronal responses, this problem has naturally received theoretical attention \cite{lochmann2012perceptual,zhu2013visual,schwabe2006feedback}. However, current literature does not describe this phenomenon as structurally integral to the neural architecture but rather either highlight a different set of features or the contextual modulations are explicitly written in an ad hoc fashion. Our aim is to develop a model which displays this phenomenon structurally, as a direct consequence of the neural architecture. In our proposed models, multiple length scales emerge naturally without any fine tuning of the system's parameters. This leads to length-tuning curves similar to the ones measured in Kapadia \textit{et al.} over the entire range (Fig \ref{fig1}) \cite{kapadia1999dynamics}.

The findings of Kapadia \textit{et al.} demonstrate that receptive fields in V1 are not constant but instead grow and shrink, seemingly beyond naive anatomical parameters, according to stimulus contrast. The ``computation'' being carried out is not fixed but is itself a function of the input. Let us examine this distinction carefully. There are numerous operations in image processing, such as Gaussian blurs or other {\em convolutional kernels}, whose spatial range is fixed. It is very natural to imagine neural circuitry having actual physical connections corresponding to the nonzero elements of a convolutional kernel, and in fact a fair amount of effort has been expended trying to identify actual synapses corresponding to such elements \cite{olshausen1996emergence,reid1995specificity}. There are, however, other image-processing operations, such as floodfill (the "paint bucket'') whose spatial extent is entirely dependent on the input; the problem of "binding'' of perceptual elements is usually thought about in this way, and mechanisms posited to underlie such propagation dynamics include synchronization of oscillations acting in a vaguely paint-bucket-like way \cite{rosenblatt1961principles,von1999and,lee2003hierarchical}. This dichotomy is artificial because these are only the two extremes of a potentially continuous range. While the responses of neurons in V1 superficially appear to be convolutional kernels, their strong dependence on input characteristics, particularly the size of the receptive field, demonstrates a more complex logic in which spatial extent is determined by specific characteristics of the input. What is the circuitry underlying this logic?

Neurons in the primary visual cortex are laterally connected to other neurons on the cortical surface and derive input from them. Experiments have shown that the spatial extent {\em on the cortical surface} from which neurons derive input from other neurons through such lateral interactions varies with the contrast of the stimulus \cite{nauhaus2009stimulus}. In the absence of stimulus contrast, spike-triggered traveling waves of activity propagate over large areas of cortex. As contrast is increased, the waves become weaker in amplitude and travel over increasingly small distances. These experiments suggest that the change in spatial summation area with increasing stimulus contrast may be consistent with the change in the decay constants of the traveling wave activity. However, no extant experiment directly links changes in summation in visual space to changes in integration on the cortical surface, and no explicit model of neural architecture has been shown to simultaneously account for, and thus connect, the input-dependence of spatial summation and lateral integration in V1. The latter one is our aim, and a crucial clue will come from the input-dependence of latencies.

Recently, a critically-balanced network model of cortex was proposed to explain the contrast dependence of functional connectivity \cite{yan2012input}. It was shown that in the absence of input, the model exhibits wave-like activity with an infinitely-long ranged susceptibility, while in the presence of input, perturbed network activity decays exponentially with an attenuation constant that increases with the strength of the input. These results are in direct agreement with Nauhaus \textit{et al.} \cite{nauhaus2009stimulus}.

We will now demonstrate that a similar model also leads to adaptive scales of spatial integration in visual space. Our model makes two key assumptions. The first is a local, not just global, balance of excitation and inhibition across the entire network; all eigenmodes of the network are associated with purely imaginary eigenvalues. It has been shown that such a critically-balanced configuration can be achieved by simulating a network of neurons with connections evolving under an anti-Hebbian rule \cite{magnasco2009self}. The second key assumption is that all interactions in the network are described by the connectivity matrix; nonlinearities do not couple distinct neurons in the network. 

There are a number of examples of dynamical criticality in neuroscience, including experimental studies in motor cortex \cite{churchland2012}, theoretical \cite{seung1998continuous} and experimental studies \cite{seung2000stability} of line attractors in oculomotor control, line attractors in decision making \cite{machens2005}, Hopf bifurcation in the auditory periphery \cite{choe1998model} and olfactory system \cite{freeman2005metastability}, and theoretical work on regulated criticality \cite{bienenstock1998regulated}. More recently, Solovey \textit{et al.} \cite{solovey2015loss} performed stability analysis of high-density electrocorticography recordings covering an entire cerebral hemisphere in monkeys during reversible loss of consciousness. Performing a moving vector autoregressive analysis of the activity, they observed that the eigenvalues crowd near the critical line. During loss of consciousness, the numbers of eigenmodes at the edge of instability decrease smoothly but drift back to the critical line during recovery of consciousness.

We also examine the dynamics of the system and show that its activity exponentially decays to a limit cycle over multiple timescales, which depend on the strength of the input. Specifically, we find that the temporal exponential decay constants increase with increasing input strength. This result agrees with single-neuron studies which have found that response latencies in V1 decrease with increasing stimulus contrast \cite{carandini1994summation,gawne1996latency,kapadia1999dynamics,albrecht2002visual}. We now turn to describing our model.

\section*{Methods}

Let $x\in\mathbb{C}^{N}$ be the activity vector for a network of
neurons which evolve in time according to the normal form equation:
\begin{equation}
\dot{x}_{i}=\underset{j}{\sum}A_{ij}x_{j}-|x_{i}|^{2}x_{i}+I_{i}(t)\label{eq:criticalNet}
\end{equation}
In this model, originally proposed by Yan and Magnasco \cite{yan2012input}, neurons interact
with one another through a skew-symmetric connectivity matrix $A$.
The cubic-nonlinear term in the model is purely local and does
not couple the activity states of distinct neurons, while the external input $I(t)\in\mathbb{C}^{N}$
to the system may depend on time and have a complex
spatial pattern.

\begin{figure}[bth]
	
	\begin{centering}
		\includegraphics[scale=0.2]{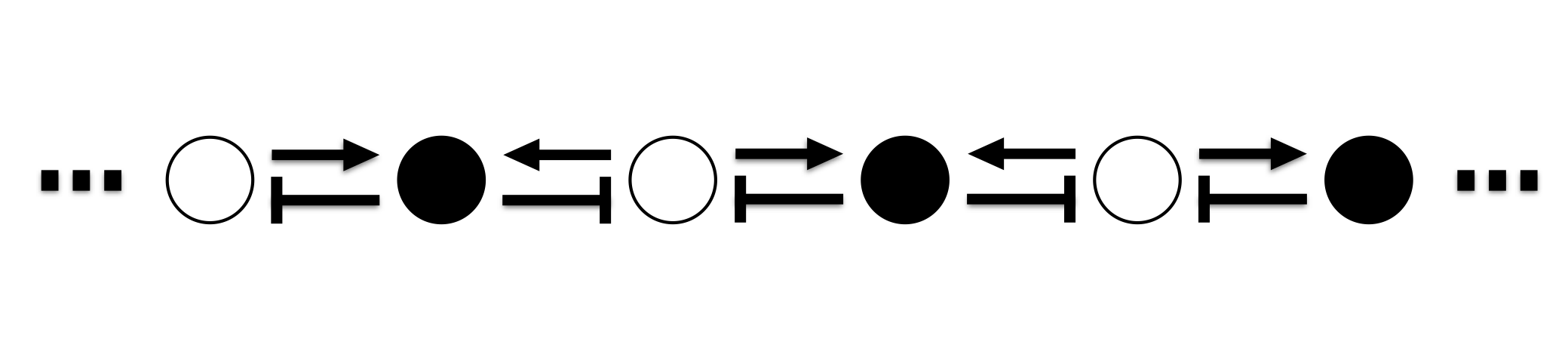}
		\par\end{centering}
	
	\begin{centering}
		\caption{\label{fig2}\textbf{Simplest connectivity matrix
				$A$}. A finite line of excitatory and inhibitory neurons. White nodes represent excitatory neurons. Black nodes are inhibitory. All connections have strength of the same magnitude. }
		\par\end{centering}
\end{figure}

The original model considered a 2-D checkerboard topology of excitatory and inhibitory neurons. For theoretical simplicity and computational ease, we will instead consider a 1-D checkerboard
layout of excitatory and inhibitory neurons which interact through equal strength,
nearest neighbor connections (Fig \ref{fig2}).
In this case, $A_{ij}=(-1)^{j}s(\delta_{i,j+1}+\delta_{i,j-1})$, where
$i,j=0,1,...,N-1$ and $s$ is the synaptic strength. Boundary conditions are such that the activity terminates to 0 outside of the finite network.

We are specifically interested in the time-asymptotic response of
the system, but explicitly integrating the stiff, high-dimensional
ODE in (\ref{eq:criticalNet}) is difficult. Fortunately, we can bypass
numerical integration methods by assuming periodic input of the form
$I(t)=Fe^{i\omega t}$, where $F\in\mathbb{C}^{N}$ and
look for solutions $X(t)=Ze^{i\omega t}$, where $Z\in\mathbb{C}^{N}$.
Substituting these into (1), we find that: 

\begin{equation}
0=(A-i\omega)Z-|Z|^{2}Z+F \label{hopfReduced}
\end{equation}

And define $g(Z)$ to be equal to the right hand side of (\ref{hopfReduced}).

The solution of (\ref{hopfReduced}) can be found numerically by using the multivariable
Newton-Raphson method in $\mathbb{C}^N$:

\begin{equation}
\widetilde{Z}\rightarrow\widetilde{Z}-J(\widetilde{Z})^{-1}\widetilde{g}(Z)\label{CritFixedPoint}
\end{equation}

where $\widetilde{Z}$ and $\widetilde{g}$  are the concatenations of the real and imaginary parts of $Z$ and $g$, respectively. $J$ is the Jacobian of $\widetilde{g}$ with respect to $$J_{ij}(z)=\frac{{\partial}\widetilde{g}_i}{{\partial}\widetilde{z}_j}$$

\section*{Results}

To test how the response of a single neuron in the network varies
with both the strength and length of the input, we select a \textit{center neuron} at index c
and then calculate, for a range of input strengths, the response of the neuron as a function of input length around it. Formally,
for each input strength level $B\in\mathbb{R}$, we solve (\ref{CritFixedPoint}) for:
\begin{equation}
F_{k}(B,l)=\begin{cases}
Bv_{k} & if\ k\in[c-l,c+l]\\
0 & otherwise
\end{cases}
\end{equation}
where $k=0,...,N-1$, {$v\in\mathbb{C}^{N}$}
describes the spatial shape of the input, and  \emph{$2l+1$} is the
length of the input in number of neurons. The response of the center neuron is taken as
the modulus of $Z_{c}$,  and we focus on the case
where $\omega$ is an eigenfrequency of $A$ and $v$ the corresponding
eigenvector. 
\begin{figure}[bth]
	
	\begin{centering}
		\includegraphics[scale=0.41]{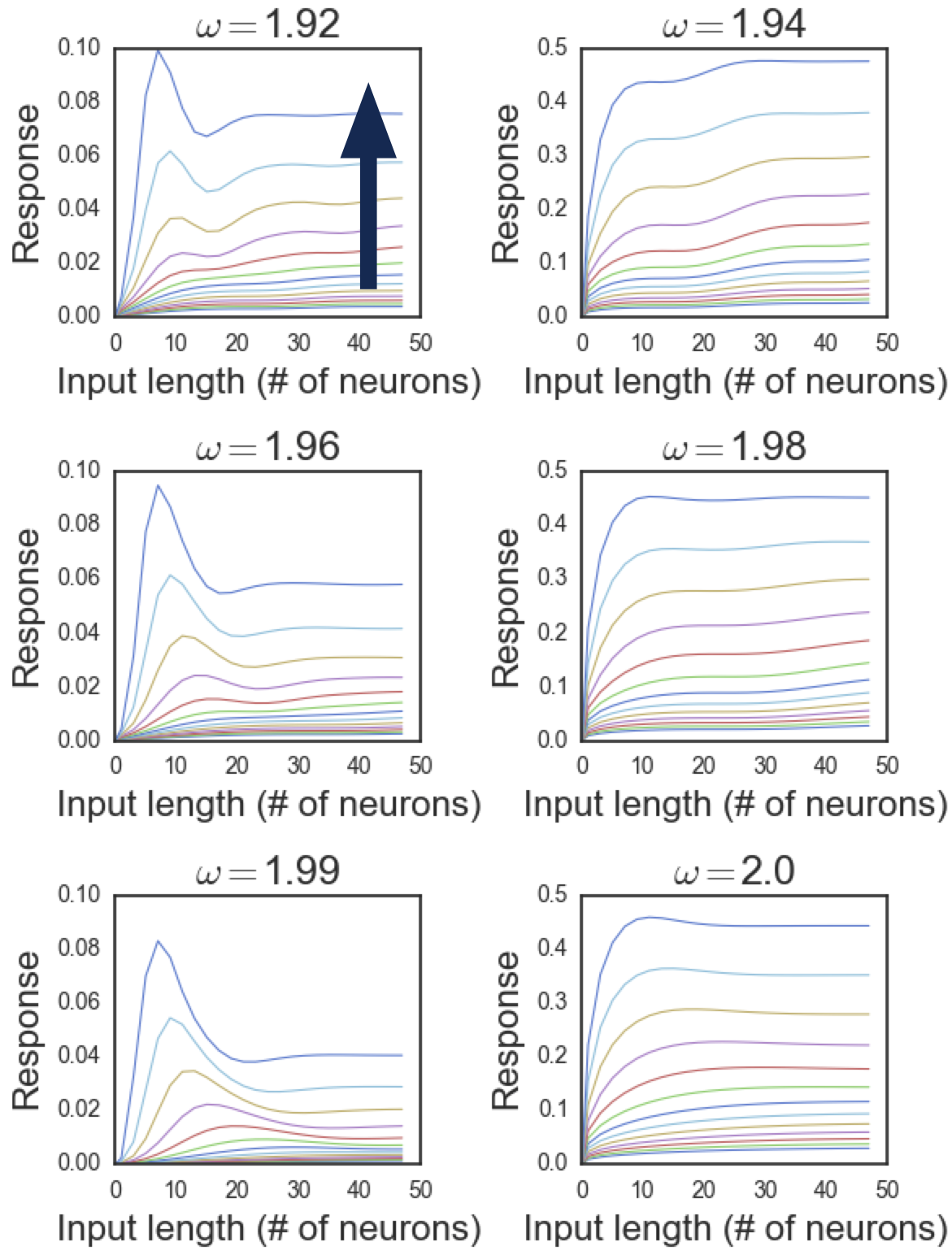}
		\par\end{centering}
	
	\centering{}\caption{\label{fig3}\textbf{Length-response curves for different eigenfrequencies. }In each panel, which corresponds to a different eigenfrequency of $A$, we plot the response of a neuron (with index N/2) as a function of input length for a group of exponentially distributed input strengths. The blue arrow in the first plot indicates the direction of increasing input strength. The length of the input is recorded
		in number of neurons and the response is taken as the modulus of the amplitude of the time-asymptotic stable limit cycle.}
\end{figure}

The results for a 1-D checkerboard network of 64 neurons is shown in Fig \ref{fig3}.
Here we fix a center neuron and sweep across a small range of eigenfrequencies $\omega$ of $A$. The curves from bottom to top correspond to an ascending order of base-2 exponentially distributed input
strengths $C=2^{i}$. For all eigenfrequencies, the peak of the response
curves shift towards larger input lengths as the input strength decreases.
In fact, for very weak input, the response curves rise monotonically
over the entire range of input lengths without ever reaching a
maximum in this finite network. This is in contrast to the response curves corresponding to strong input, which always reach a maximum but, depending on the eigenfrequency,
exhibit varying degrees of response suppression beyond the maximum. This is consistent with variability of response suppression in primary visual cortex
studies \cite{kapadia1999dynamics,sceniak1999contrast}. In Fig \ref{fig3},
eigenfrequencies $\omega=1.92, 1.96, 1.99$ show the greatest amount
of suppression while the others display little to none.

\begin{figure}[bth]
	
	\begin{centering}
		\includegraphics[scale=0.33]{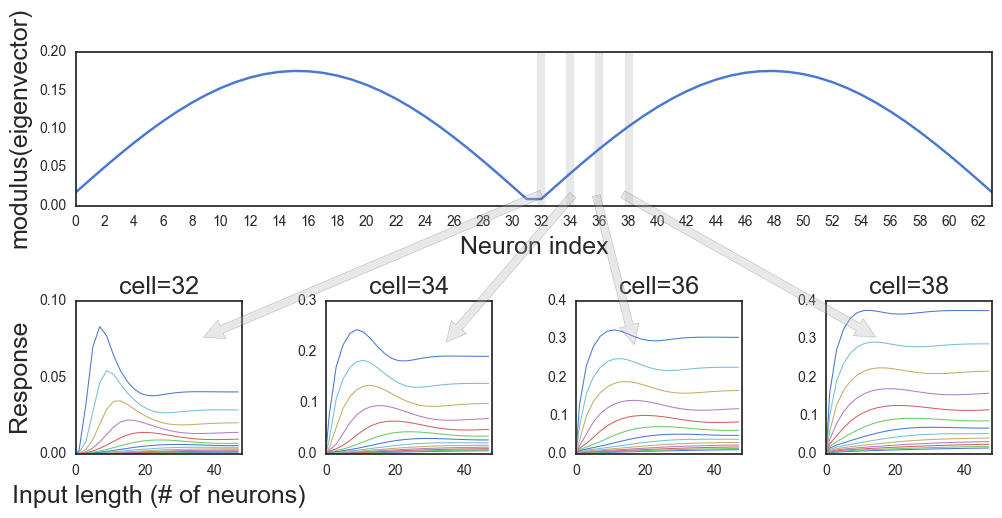}
		\par\end{centering}
	
	\begin{centering}
		\caption{\label{fig4}\textbf{Length-response curves of different
				neurons. }The top plot depicts the modulus of the eigenvector corresponding
			to eigenfrequency $\omega=1.92$. The 4 panels below are plots of
			the length-response curves for 4 different neurons in the network.
			The position of the neurons relative to the shape of the eigenvector
			are noted with the gray bars and arrows.}
		\par\end{centering}
	
\end{figure}

To understand why certain eigenfrequencies lead to suppression, we
fix the eigenfrequency to be $\omega=1.92$ and examine the response
curves of different center neurons. The response of four center neurons
(labeled by network position) and the modulus of the eigenfrequency's
corresponding eigenvector are plotted in Fig \ref{fig4}.
The center neurons closest to the zeros of the eigenvector experience
the strongest suppression for long line lengths. Neuron 38 closer to the
peak of the eigenvector's modulus experiences almost zero suppression.
This generally holds for all eigenvectors and neurons in the network
as all eigenvectors are periodic in their components with an eigenvalue-dependent spatial frequency.

To strengthen the connection between model and neurophysiology, one can consider a critically-balanced network with an odd number of neurons so that 0 is now an eigenfrequency of the system. In our model, input associated with the 0-eigenmode represents direct current input to the system which is what neurophysiologists utilize in experiments; the visual input is not flashed \cite{kapadia1999dynamics,sceniak1999contrast}. Contrary to the even case, long range connections must be added on top of the nearest neighbor connectivity in order to recover periodic eigenvectors and hence suppression past the response curves maximums. 

Next, we show that the network not only selectively integrates
input as a function of input strength but also operates on multiple time scales which flexibly adapt to the input. This behavior is not surprising given that in the case of a single critical Hopf oscillator, the half width of the resonance, the frequency range for which the oscillator's response falls by a half, is proportional to the forcing strength of the input, $\Gamma\propto{F^{\frac{2}{3}}}$ where $\Gamma$ is the half-width $F$ the input strength \cite{eguiluz2000essential}. Thus, decay constants in the case of a single critical oscillator should grow with the input forcing strength as $F^{\frac{2}{3}}$.

\begin{figure}[bth]
	
	\begin{centering}
		\includegraphics[scale=0.32]{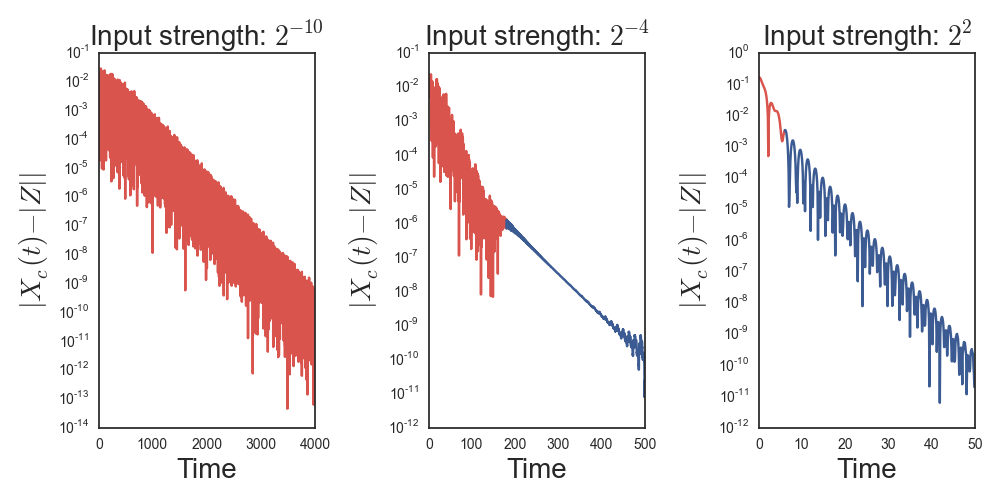}
		\par\end{centering}
	
	\begin{centering}
		\caption{\label{fig5}\textbf{Input-strength-dependent timescales}. For neuron $i$ in the network, we plot $\log\left(\left||x_{i}(t)|-\left|Z_{i}\right|\right|\right)$ as a function of time for 3 different input strengths. Linear regions correspond to exponential decay. In the presence of weak input, a fast decay regime (red) guides the dynamics towards a stable limit cycle. For the intermediate input strength, a new, distinct slow decay regime appears (blue), which becomes dominant for strong input}
		\par\end{centering}
\end{figure}

Assuming input $Fe^{i{\omega}t}$, as described above, the network activity $x(t)$, given by (\ref{eq:criticalNet}), decays exponentially in time to a stable limit cycle, $X(t)=Ze^{i\omega t}$. This implies that for any neuron $i$ in the network, $|x_{i}(t)|=e^{-bt}f(t)+|Z_{i}|$ during the approach to the limit cycle. We therefore plot $\log\left(\left||x_{i}(t)|-\left|Z_{i}\right|\right|\right)$ over the transient decay period  and estimate the slope of the linear regimes. We do this for a nearly network size input length (input length=29, $N=32$) and a range of exponentially distributed input strengths. In Fig \ref{fig5}, we plot representative transient periods of a single neuron corresponding to 3 input strengths: $2^{-10}$, $2^{-4}$, and $2^{2}$. For weak input there is a fast single exponential decay regime (red) that determines the system's approach to the stable limit cycle. As we increase the input, however, the transient period displays two exponential decay regimes: the fast decay regime (red) which was observed in the presence of weak input and a new slow decay regime (blue) immediately preceding the stable limit cycle. For very large input strength, the slow decay regime becomes dominant. The multiple decay regimes is a surprising result which doesn't appear in the case of a single critical Hopf oscillator. 

\begin{figure}[bth]
	
	\begin{centering}
		\includegraphics[scale=0.42]{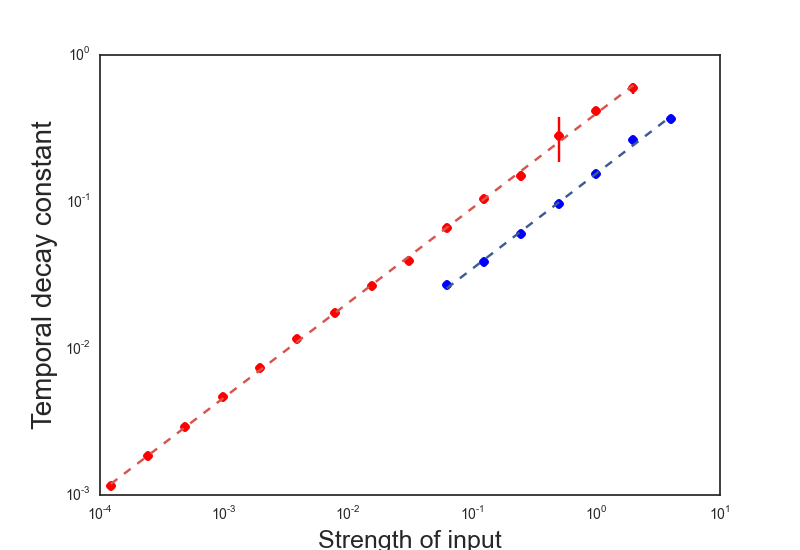}
		\par\end{centering}
	
	\begin{centering}
		\caption{\label{fig6}\textbf{Input-strength dependence of exponential decay constants}. The temporal exponential decay constants for a range of input strengths is depicted above. A fast decay regime (red circles)  is accompanied by a slow decay regime (blue circles) at large input strengths. Each decay regime is separately approximated by a least squares line (dashed lines) in log-log space.}
		\par\end{centering}
\end{figure}

We estimate the exponential decay constants as a function of input strength and plot them on a log-log scale in Fig \ref{fig6}. The red circles correspond to the fast decay regime, while the blue circles correspond to the slow decay regime, which becomes prominent for large forcings. We separately fit both the slow and fast decay regimes with a best fit line.  Unsurprisingly, the slopes of the lines are equal and approximately $\frac{2}{3}$. Thus, the decay constants grow with the input as $\propto$ $F^{\frac{2}{3}}$, where F is the input strength. This implies that the system operates on multiple timescales dynamically switching from one to another depending on the magnitude of the forcing. Larger forcings lead to faster network responses. 

In this paper, we consider a line of excitatory and inhibitory neurons, but our results hold equally well for a ring of neurons with periodic boundary conditions and appropriately chosen long range connections. Ring networks have extensively been studied as a model of orientation selectivity in V1 \cite{bressloff2000dynamical,bressloff2001geometric,shriki2003rate,ermentrout1998neural,hansel1997modeling,yishai1995theory,dayan2001theoretical}. In agreement with recent findings \cite{rubinr2015suprlinear}, the critically-balanced ring network exhibits surround suppression in orientation space when long range connections are added on top of nearest neighbor connectivity.

\section*{Conclusion}

We have shown that a simple dynamical system poised at the onset of instability exhibits an input-strength-dependent scale of integration of the system's input and input-strength-dependent response latencies. This finding strongly complements our previous results showing that a similar nonlinear process with fixed, nearest neighbor network connectivity leads to input-dependent functional connectivity. This system is thus the first proposed mechanism that can account for contrast dependence of spatial summation, functional connectivity, and response latencies. In this framework, these three characteristic properties of signal processing in V1 are intrinsically linked to one another.

%
%
%

\end{document}